\documentclass[10pt,a4paper]{article}

\usepackage[authoryear]{natbib}

\usepackage{hyperref}

\usepackage{graphicx}

\title{Economics 2.0: The Natural Step towards A Self-Regulating, Participatory Market Society}

\author{Dirk Helbing$^1$\\
$^1$ETH Zurich, Swiss Federal Institute of Technology,\\ Department of Humanities, Social and Political Sciences,\\ Clausiusstrasse 50, 8092 Zurich, Switzerland}

\begin{document}

\maketitle




\begin{abstract}

Despite all our great advances in science, technology and financial innovations, many societies today are struggling with a financial, economic and public spending crisis, over-regulation, and mass unemployment, as well as lack of sustainability and innovation. Can we still rely on conventional economic thinking or do we need a new approach? Is our economic system undergoing a fundamental transformation? Are our theories still doing a good job with just a few exceptions, or do they work only for ``good weather'' but not for ``market storms''? Can we fix existing theories by adapting them a bit, or do we need a fundamentally different approach? These are the kind of questions that will be addressed in this paper.

I argue that, as the complexity of socio-economic systems increases, networked decision-making and bottom-up self-regulation will be more and more important features. It will be explained why, besides the ``homo economicus'' with strictly self-regarding preferences, natural selection has also created a ``homo socialis'' with other-regarding preferences. While the ``homo economicus'' optimizes the own prospects in separation, the decisions of the ``homo socialis'' are self-determined, but interconnected, a fact that may be characterized by the term ``networked minds''. Notably, the ``homo socialis'' manages to earn higher payoffs than the ``homo economicus''.

I show that the ``homo economicus'' and the ``homo socialis'' imply a different kind of dynamics and distinct aggregate outcomes. Therefore, next to the traditional economics for the ``homo economicus'' (``economics 1.0''), a complementary theory must be developed for the ``homo socialis''. This economic theory might be called ``economics 2.0'' or ``socionomics''. The names are justified, because the Web 2.0 is currently promoting a transition to a new market organization, which benefits from social media platforms and could be characterized as ``participatory market society''. 

To thrive, the ``homo socialis'' requires suitable institutional settings such a particular kinds of reputation systems, which will be sketched in this paper. I also propose a new kind of money, so-called ``qualified money'', which may overcome some of the problems of our current financial system.

In summary, I discuss the economic literature from a new perspective and argue that this offers the basis for a different theoretical framework. This opens the door for a new economic thinking and a novel research field, which focuses on the effects, implications, and institutional requirements for global-scale network interactions and highly interdependent decisions.

\end{abstract}











\section{Introduction}

In the past decades, our world has changed more quickly than ever. Globalization and technological revolutions have created a world with many systemic interdependencies, high connectivities, and great complexity. The nature and intensities of many 21st century problems are an immediate consequence of this \citep{HelbingGloballyNetworkedRisks}. However, we lack a ``Global Systems Science'' to understand the world we have created, and many of our institutions are conceptually outdated.

In fact, the intellectual framework of the current institutions of our socio-economic system is around 300 years old for example the work of Adam Smith (1776) on the ``Nature and Causes of the Wealth of Nations''. Many attribute to him the idea of an ``invisible hand'' that coordinates individual interests by self-organization. Specifically, it is often believed that optimally self-regarding behavior, as assumed by the standard microeconomic paradigm of the ``homo economicus'', would create maximum social welfare. This idea goes back to the so-called Fable of Bees \citep{MannevilleFableOfBees} and is the basis of neoclassical economics.\footnote{Note, however, that the bees belonging to a hive have basically identical genes, such that their interests are expected to be aligned from the perspective of evolutionary competition.} Even  today, many policies are founded on it, including those that demand homogeneous and free global markets.\footnote{This includes, for example, the World Bank and the International Monetary Fund.}

In fact, it can be shown that the ``invisible hand'' works well, if all conditions assumed by the First Theory of Welfare Economics are fulfilled \citep{Arrow54,DebreuTheoryOfValue}. However, under many realistic conditions at least one of the assumptions is not satisfied so that the welfare-optimal outcome is not guaranteed. This is, for example, the case if individuals interact in networks rather than through one homogeneous market, or if externalities or transaction costs matter \citep{Coase1960,RocaHelbingPercolationChapterInSocialSelfOrganization}. Such conditions can lead to ``market failures''. The same applies to information asymmetries or cases with powerful monopolists or oligopolists \citep{akerlof1970market,mas1995microeconomic}.

It is therefore understandable that Thomas Hobbes (1651) pointed out, a selfish species would require regulation. In order to create socio-economic order, he proposed that a ``Leviathan''--- a  powerful state to tame the ``selfish beast''.\footnote{The famous quote ``homo hominis lupus'' compares humans with wolves.} Even today, this constitutes the intellectual framework of the need of top-down regulation, as it also shows up in Keynesianism. So, is the idea of a decentrally self-organizing economy flawed? On the contrary, we have just not learned, how self-regulation, i.e. bottom-up regulation, can work.\footnote{Here, the term ``self-organization'' is used for the emergence of collective behavior or properties, which may be desireable or not. ``Self-regulation'' is used for a set of rules that supports the adaptive self-organization of a favorable outcome such as the convergence to a system-optimal state.}

The idea of self-organization is still extremely appealing and timely \citep{holldobler}. Ants, for example, work in an entirely self-organized and largely decentralized way without a hierarchical system of command. The same applies to social animals like termites, flocks of birds, or schools of fish. The ecological and immune systems also function in a decentralized and highly efficient way, due to the evolutionary principles of mutation and selection. Social norms are success principles building on decentralized mechanisms as well. I will argue that we can learn from principles like these to make the ``invisible hand'' work, such that individual and social benefits can be simultaneoulsy reached.

It is a great scientific challenge to find principles of self-regulation that work under less optimistic conditions than those assumed by the Theorems of Welfare Economics. In particular, they should work in case of social dilemma situations, where cooperative behavior would be beneficial for everyone, but where exploiting others can create even higher individual payoffs. I would like to point out that social dilemma situations are expected to be common in socio-economic systems, since opportunities to reach higher benefits by overpriced low-quality products or services or by misleading information exist in many economic exchange situations. For example, the value of a product might be lower, or the risk of a financial derivative might be higher than claimed. Under such conditions, cooperation can easily erode, leading to ``tragedies of the commons''. Well-known examples are environmental pollution, overfishing, global warming, free-riding, and tax evasion. The breakdown of trust and cooperation may also be seen as the main cause of the financial meltdown of 2008, i.e. it could also be interpreted as ``tragedy of the commons'' \citep{HelbingGloballyNetworkedRisks}.

Therefore, to be successful, societies must be able to deal with ``social dilemma situations''. Unfortunately, the self-regarding behavior of a ``homo economicus'' (as reflected by a ``best response rule'' in our later discussion), is destined to lead to ``tragedies of the commons''. The current solution to this is to introduce taxes and regulations that change the nature of the interactions and eliminate the occurence of social dilemma situations \citep{HelbingPD}. But is this  approach effective, and is it the best solution? The de facto failure of the international carbon tax raises doubts that it does. So, is a totally different approach required?

It is often assumed that other-regarding behaviour is at the cost of personal disadvantage. If so, due to the merciless forces of natural selection, nothing other than a self-regarding ``homo economicus'' should exist. In Sec. \ref{homosocialis} I will present computer simulations that test this assumption. Surprisingly, biological evolution can also be shown to produce a ``homo socialis'' with other-regarding preferences, which may overcome ``tragedies of the commons''.\footnote{Note that Adam Smith himself believed more in a competitive, but other-regarding human being, than in a maximising self-regarding ``homo economicus''.  In his book ``The Theory of Moral Sentiments'' (1759) he writes: ``How ever selfish man may be supposed, there are evidently some principles in his nature, which interest him in the fortune of others, and render their happiness necessary to him, though he derives nothing from it. Of this kind is pity or compassion, the emotion which we feel for the misery of others, when we either see it, or are made to conceive it in a very lively manner. That we often derive sorrow from the sorrow of others, is a matter of fact too obvious to require any instances to prove it; for this sentiment, like all the other original passions of human nature, is by no means confined to the virtuous and humane, though they perhaps may feel it with the most exquisite sensibility. The greatest ruffian, the most hardened violator of the laws of society, is not altogether without.'' Considering this, one might conclude that Adam Smith's concept of humans and how an economy would self-organize has still not been fully formalized by mainstream economics or practically implemented.} 

However, this type of actor cannot thrive well in an institutional framework created for a ``homo economicus''. The ``homo socialis'' requires different institutions. So, what institutions do we need in the 21st century?

In the past, societies have invented various institutions to overcome ``tragedies of the commons'', for example, genetic favoritism, direct reciprocity (``I help you, if you help me''), or punitive institutions (such as police, courts, prisons and regulatory authorities). However, all of these mechanisms can lead to undesirable side effects such as civil wars or corruption. Our current system is built on punitive institutions, but such top-down regulation is very costly and creates many inefficiencies, including negative impacts on innovation \citep{Inno,Inno2,Inno3,Helbing05b}.\footnote{Some of the leading industrial countries now have debts around 100\% of the GDP, which suggests that systems with many institutions for top-down regulation are very expensive and may not be sustainable on the long run.}

In Sec. \ref{trafficlight}, I will explain that, as systems get more and more complex, top-down regulation can no longer achieve efficient and satisfactory solutions. Instead, we can build on self-regulation, such as found in ecological, social and immune systems. I will illustrate the advantage of this approach by the example of urban traffic light control and will argue that a global reputation platform could become a suitable institution for our globalized world, the ``global village'' that we have created (see Sec. \ref{ReputationSystems}). This proposed approach supports trusted exchange, and increases opportunities for the participation of citizens in social, economic and political life. Furthermore, I will present the idea of a new, reputation-based kind of money, which I call ``qualified money'' (see Sec. \ref{qualifiedMoney}). These new institutions would promote an ``economy 2.0'', which will be a ``participatory market society'', and help to overcome the current gap between social and economic engagement (see Sec. \ref{ParticipatoryMarketSociety}).

My overall conclusion is that a change from an agent-oriented to an interaction-oriented view by decision-makers offers a better understanding of complex socio-economic systems, and leads to new solutions for long-standing problems. I will cite examples from the technology and business world, that demonstrate that the transition to a ``participatory market society'', is already occuring. This is fueled by the digital revolution, particularly the Web 2.0 and social media platforms. If the right political decisions are taken pro-actively, this scenario can unleash the potentials of many ``networked minds'' and lead to an age of creativity and sustainable prosperity.

\section{A Paradigm Shift in Economic Thinking?}

Like mathematics and physics, economics is proud of having an axiomatic foundation, and rightly so. In microeconomics, which focuses on the decision-making of actors such as individuals or firms, the paradigm of the self-regarding, optimizing agent prevails \citep{Neumann44}. It also forms the basis of the theory of markets \citep{Arrow54}. In essence, in macroeconomics there are basically two major competing schools: the Neoliberals who believe in free markets \citep{WalrasElements,Marshall1890,Lucas79} and the Keynesians who call for market regulation by politics \citep{KeynesTheGeneral, Mankiw91}. These theories still dominate much of public economic thinking and policy-making around the world. However, there are a number of unresolved issues \citep{Krugman09, Colander2009, Kirman2010, HaldaneMaySystemicRisk2011, LuxWesterhoffEconomicsCrisis,  JohnsonLux2011, DrawingBoardMacroeconomics, HelbingRethinkingEconomics, HelbingBaliettiChallenges}:




\begin{enumerate}

\item The two macroscopic schools have incompatible views of economics \citep{GreenwaldStiglitzKeynesian, Keen13}.

\item A consistent theoretical link between micro- and macroeconomics (which goes beyond the simplifying view of representative agent modeling) is lacking \citep{KirmanRepresentativeIndividual}.

\item Empirical and experimental findings are challenging some of the foundational assumptions of economics \citep{Kahneman79, Kahneman84, Kahneman96, Hoffman96, Hartley97, Selten98,  Tversky92, Fehr99, Fehr03, Ariely08}.

\item The prevailing microeconomic view of decision-making is not well compatible and integrated within the body of knowledge collected in anthropology, social psychology and sociology \citep{lindenberg,GintisBoundsOfReason,McFadden}.

\item The financial crisis has raised serious doubts that the most established economic theories can sufficiently  describe financial meltdowns of markets and their impact on economies and societies \citep{Colander2009, Kirman2010, Keen11, gersbach, tirole, Keen13}.

\end{enumerate}


Are we perhaps facing economic problems, because our theoretical picture of economies does not fit reality well? To answer this, scientists need to question established knowledge. In fact, many new approaches have been proposed in the past decades. Due to spatial limitations I can mention only a few. For example, I would refer to the work of Brian Arthur on innovation (1989, 1999), of Reinhard Selten in experimental economics (1998), of Joseph Stiglitz on economic downturns (2011), of Paul Krugman on the role of geography (1990), of Daniel Kahnemann and Amos Tversky on risk perception (1979, 1984, 1992, 1996), of George Akerlof and Robert Shiller on herding behavior (``animal spirits'', see \cite{AkerlofAnimalSpirits}), of Gerd Gigerenzer (1999) on heuristics, of Bruno Frey (2008) on happiness, of Ernst Fehr on fairness (1999, 2003), of Herbert Gintis on evolutionary game theory (2009), of Ulrich Witt on evolutionary economics (1998), and of Alan Kirman and Mauro Gallegati on the role of heterogeneity and non-representative agents in economics \citep{Kirman06, BeyondtheRepresentativeAgent,HeterogeneousInteractingAgents,EmergentMacroeconomics}. I would also mention work presenting a complexity science perspective on economic systems \citep{KrugmanSelfOrganizing1996,DayComplexEconomicDynamics1999,AuyangComplexSystemTheories1999,CopingComplexity2009,KirmanComplexEconomics,Brock97,Lux99, HelbingBaliettiChallenges,Keen13} or an econophysics perspective \citep{AokiReconstructingMacroeconomics, BouchaudFinancialRisks, SornetteWhyStockMarketsCrash, EconophysicsIntroduction2011, EconophysicsOfMarketsChatterjee, EconophysicsSociophysicsChakrabarti}. So far, however, a unifying theoretical framework of all these important findings and their derivation from first principles is largely lacking, in contrast to mainstream economics. Perhaps for this reason, many foundational economics courses still suggest that markets can be understood \emph{as if} people behaved according to the idealized assumptions of mainstream economics.\footnote{The book by \cite{UnderstandingCapitalism} appears to be an exception.} It, therefore, seems that most economic work still views man as an -- in many ways -- imperfect approximation of a ``homo economicus'', and that these imperfections would more or less average out on the macro-level.

In response to the low predictive power of many economic theories and the economic problems most countries face today, it is often argued that mainstream economic principles are idealizations rather than a faithful representation of the real world. This calls for a better understanding of the deviations between currently established theories on the one hand, and empirical or experimental evidence on the other. Perhaps it even requires a fundamentally new way of thinking.

In fact, many eminent thinkers call for a different approach. They include Nobel prize laureates and ex-presidents of central banks. The most explicit statements are probably those of Paul Krugman, who in 2009 asked: ``How did economists get it so wrong?'', and Alan Greenspan, who acknowledged in testimony before a congressional committee on October 23, 2008, that ``the whole intellectual edifice ... collapsed in the summer of last year''. For more quotes see \cite{HelbingBaliettiChallenges} and \cite{HelbingRethinkingEconomics}. In adddition, most financial traders do not seem to apply nor believe in scholarly economics. George Soros has established an \emph{Institute of New Economic Thinking}\footnote{See \url{http://ineteconomics.org}}, Edward Fullbrock and others have founded the \emph{World Economics Association} to support heterodox economics with journals such as \emph{Real-World Economics Review}.\footnote{See \url{http://www.worldeconomicsassociation.org} and \url{http://www.paecon.net/PAEReview}}

While theories in microeconomics are mostly based on the concept of the ``homo economicus'', it must also be recognized that identical macro-level phenomena might be explained by different kinds of micro-level assumptions, as much relevant information is lost through macro-level aggregation \citep{HelbingPluralisticModeling}. There are alternative micro-level theories which are worth considering \citep{GintisBoundsOfReason}. For example, rather than taking the ``homo economicus'' as a theoretical starting point, it would be equally justified to start from the ``homo socialis'', i.e. decision-makers with other-regarding preferences (see Sec.\ref{homosocialis}).

While there is empirical evidence for other-regarding behavior \citep{Gueth82,Henrich01,Fehr03,BFrey} and some theoretical implications of this have been studied \citep{Dufwenberg, Sobel}, a good theoretical foundation of the ``homo socialis'' has long been lacking. Recent theoretical progress \citep{Grund13}, however, suggests that a theory for the ``homo socialis'' can be rigorously derived from first principles, rooted in the best response rule (utility maximization) and principles of evolution (see Sec.\ref{homosocialis}). This raises the pertinent question whether such a theory might better explain  empirical and experimental evidence, and perhaps create a paradigm shift in economic thinking in due course.\footnote{Empirical evidence suggests that more than 60\% of subjects have other-regarding preferences \citep{Murphy11}, but the percentage may also depend on the socio-cultural background.}

For example, the fact that the ``homo socialis'' is influenced by the interests and preferences of others explains a mysterious, but widely observed fact in sociology, namely social influence \citep{Asch51, Asch56}. Social influence helps to understand typical characteristics of opinion formation \citep{Maes10} and herding effects (``animal spirits''), including bubbles and crashes at financial markets \citep{AkerlofAnimalSpirits, Smith88, CaginalpBubbles,Hommes05, Hommes08, SornetteWhyStockMarketsCrash}.

I would also expect that the incorpration of the theory of the ``homo socialis'' will provide a new insights and a fundamentally new understanding of social capital, power, reputation, trust as well as economic value \citep{LinSocialCapital}, as these quantities result from social network interactions. Such network interactions are characteristic for the ``homo socialis'', who is perhaps best characterized by the term ``networked minds'' (see below). I expect that a theory of the ``homo socialis'' will be able to consistently explain many ``puzzles'' in the social sciences and economics. Currently, we still know little about the ``homo socialis'', but what we do know is very interesting. Therefore, there is much exciting research to be done. I argue that the ``homo socialis'' would justify a new branch of economics that one might call ``economics 2.0'' or ``socionomics'' (see Sec. \ref{Socionomics}).\footnote{Note that the methodological approach and scientific understanding proposed in this paper has little overlap with the current, mood-focused approach by \cite{prechter}, see http://www.socionomics.net/.}
So, what exactly is the ``homo socialis'' about?




\section{The Emergence of the ``Homo Socialis''}\label{homosocialis}

Establishing a new economic thinking is not just obstructed by a lack of alternative (``heterodox'') models. If one departs from the concept of the ``homo economicus'', i.e. the strictly optimizing self-regarding agent, there are myriads of possibilities how agents and their decision-making rules may be defined. However, no alternative model stands out thus far, and a consensus on a non-homo-economicus-kind of model has not emerged. In some sense, there is too much arbitrariness in specifying such models. But this might just have changed.

\subsection{Utility-Maximizing Agents Under Evolutionary Pressure}

A recently published paper \citep{Grund13} studied whether evolution has made humans ``nasty'' (self-regarding) or ``nice'' (other-regarding)\footnote{See also the related article in the Wall Street Journal of 29.03.2013: \url{http://online.wsj.com/article/SB10001424127887324105204578384930047065520.html}.}. It analyzes social dilemma situations for the typical example of the ``Prisoner's Dilemma game'' (PD), which distinguishes two kinds of behavior: ``cooperation'' and ``defection''. The PD takes into account the risk of cooperation and the temptation to defect.

For Prisoner's Dilemma interactions, cooperative behavior is irrational for a ``homo economicus'', resulting in a ``tragedy of the commons'' with poor payoffs \citep{Hardin}. To stabilize cooperation, social mechanisms such as direct, indirect, spatial or network reciprocity, group competition, or costly punishment have been proposed, often assuming imitation of better-performing behaviors \citep{GINTISEvolutionaryGameTheory,NowakEvolutionary, SigmundExplorations,Fehr03}. These mechanisms, like taxes, effectively transform the Prisoner's Dilemma into other kinds of games that are more favorable for cooperation \citep{HelbingPD}. So far, however, other-regarding preferences\footnote{not to be confused with other-regarding behavior, which is the same as cooperation} were lacking a convincing explanation.\footnote{i.e. an explanation, which does not assume favorable but debatable mechanisms such as genetic drift or cultural selection (in the sense of a social modification of the reproduction rate)}

The above mentioned study \citep{Grund13} distinguishes individual preferences from behavior. It investigates Prisoner's Dilemma interactions in two-dimensional space on the basis of very few assumptions, each of which promotes self-regarding preferences or defection:




\begin{enumerate}

\item Agents decide according to a best-response rule that strictly maximizes their utility function, given the behaviors of their interaction partners (their neighbors).

\item The utility function considers not only the own payoff, but gives a certain weight to the payoff of their interaction partner(s). The weight is called the ``friendliness'' and set to zero for everyone at the beginning of the simulation.

\item Friendliness is a trait that is inherited (either genetically or by education) to offspring. The likelihood to have an offspring increases exclusively with the own payoff, not the utility function. The payoff is assumed to be zero, when a friendly agent is exploited by all neighbors (i.e. if they all defect). Therefore, such agents will never have any offspring.

\item The inherited friendliness value tends to be that of the parent. There is also a certain mutation rate, but it does not promote friendliness. (In the simulation results discussed here, mutations were specified such that they imply an average friendliness of 0.2, which cannot explain the typically observed value of 0.4.)

\end{enumerate}


\begin{figure}[h!]

\centerline{\includegraphics[width=0.85\textwidth]{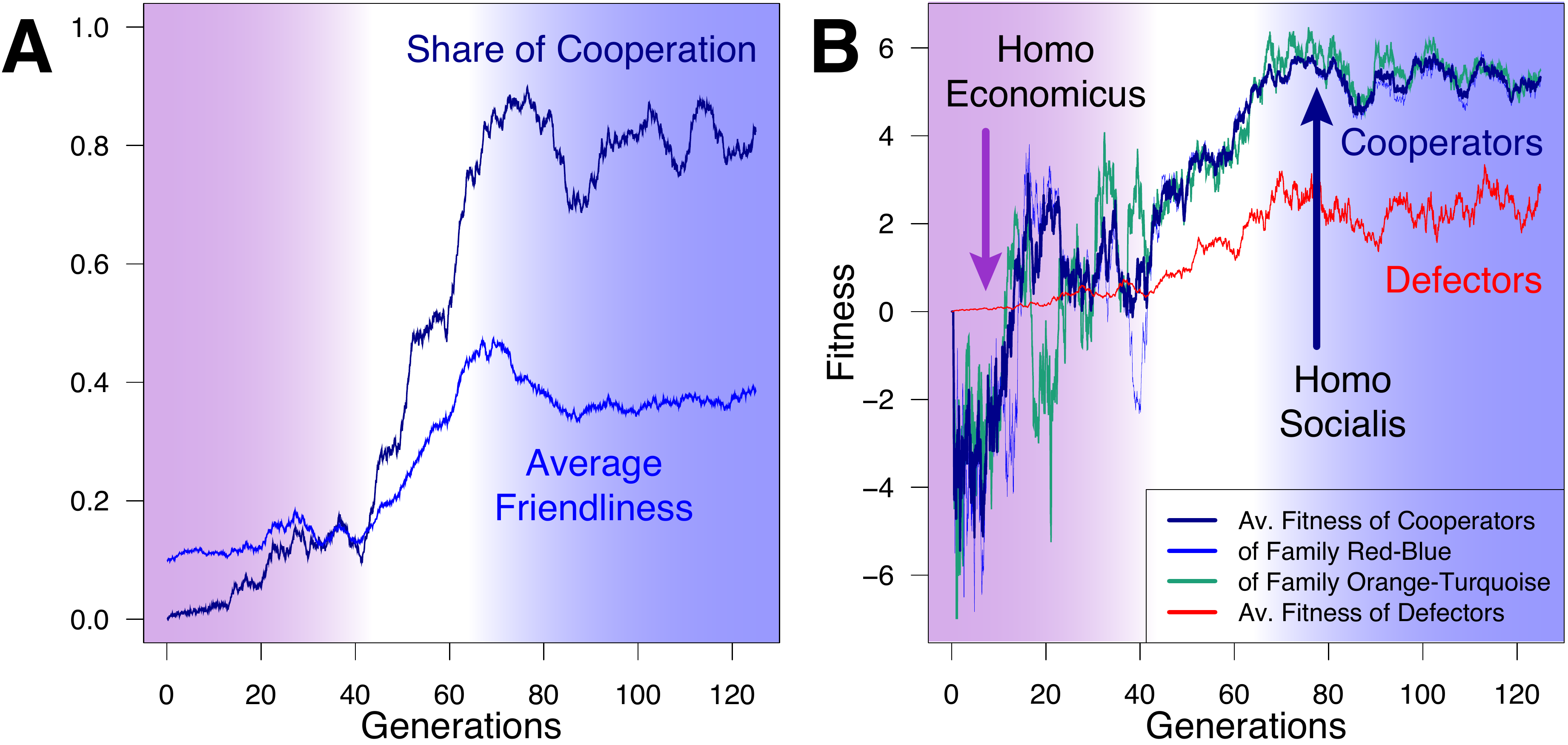}}

\caption[]{\textbf{Emergence of the ``homo socialis'' in a world initially dominated by the ``homo economicus''.}

(A) Average value of the friendliness and of the proportion of cooperating agents as a function of time. One generation corresponds to $\frac{1}{\beta}$ update time steps, where $\beta$ is the death rate.

(B) Average payoffs of cooperators and defectors as a function of time. At the beginning, defectors are more successful than cooperators, as they receive higher payoffs on average. However, after many generations, a transition from a ``homo economics'' to a ``homo socialis'' occurs. Then, the payoffs for cooperating agents (which are of the type ``homo socialis'') are higher than the payoffs of defectors. This allows the ``homo socialis'' to spread thanks to a reproduction rate that is now higher than that of the ``homo economicus''. Note that the population dynamics implies that families might die out before high levels of cooperation are reached. For details of the underlying simulations see the caption of Fig. \ref{FigCoopStrangers} and \cite{Grund13}.

}

\label{FigEmergence}

\end{figure}

\begin{figure}[h!]

\centerline{\includegraphics[width=.8\textwidth]{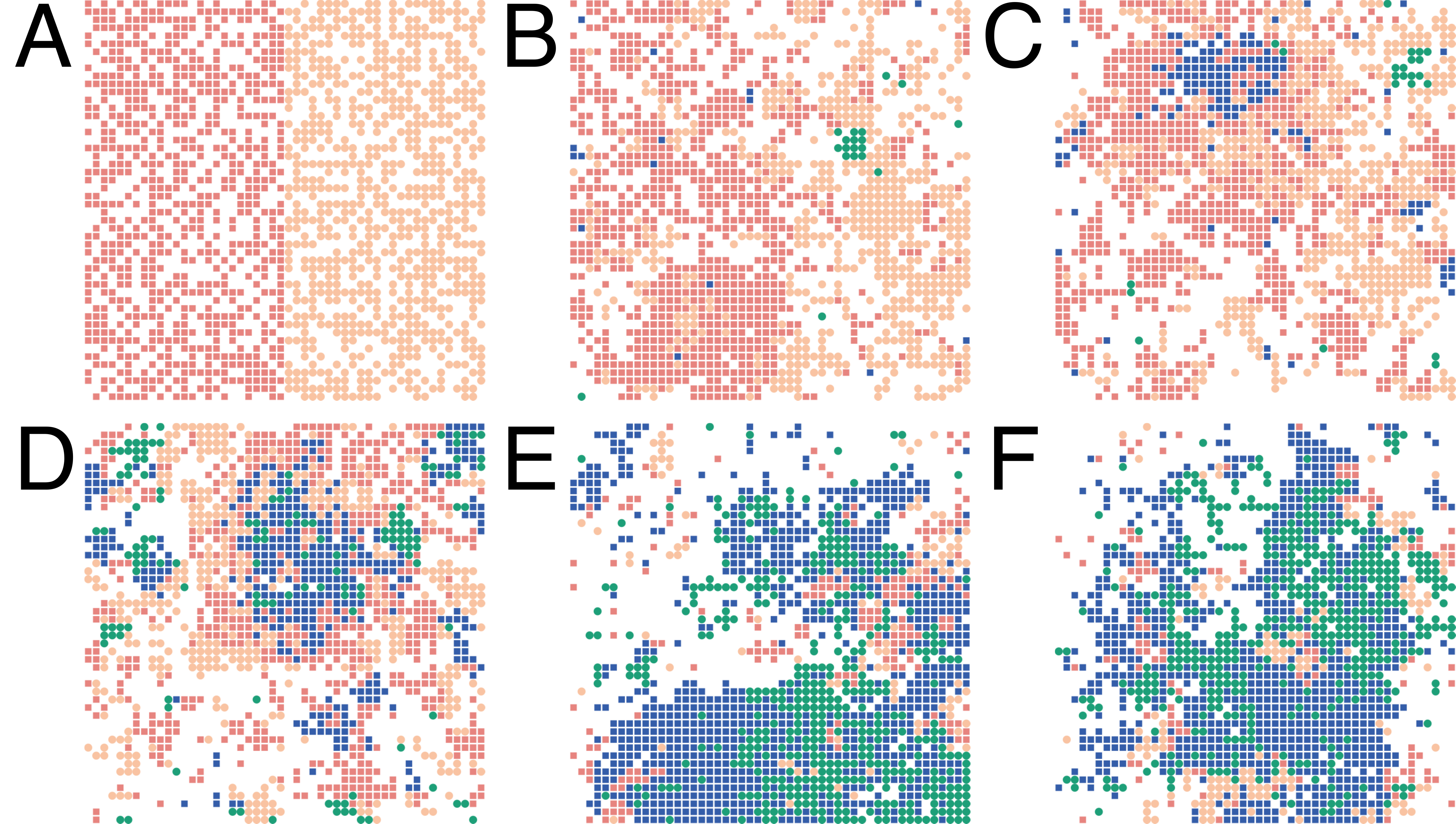}}

\caption[]{\textbf{Evolution of a ``cooperation between strangers'' (here: cooperation between two different families).}

The above figure shows snapshots (at times $t=0, 250, 500, 1000, 2000$, and $t=2500 =125$ generations) of a representative simulation run for two families (i.e. two kinds of genetically related individuals), populating $60\%$ of a $50\times 50$ spatial grid without periodic boundary conditions. For a related video see \url{http://vimeo.com/65376719}. The initial friendliness in the orange-turquoise family is 0, and in the red-blue family it is assumed to be 0.2 (but it could be chosen 0 as well, see \cite{Grund13}).  One finds the emergence of cooperation in both families (represented by blue or turquoise), which cannot occur for the ``homo economicus''. Hence, cooperation implies a ``homo socialis''. Remarkably, there are clusters in which members of different families cooperate (i.e. a ``cooperation between strangers'' evolves). Individuals play 2-person prisoner's dilemma games with all neighbors in their respective Moore neighborhoods (but similar results are expected for some other social dilemma games as well). The payoff parameters are the ``Reward'' $R = 1$ for the cooperation of both interaction partners, the ``Punishment'' $P=0$ for non-cooperative behavior on both sides, the ``Temptation'' $T = 1.1$ for a defector exploiting a cooperator, and the ``Sucker's Payoff'' $S = -1$ for an exploited cooperator. Each agent's utility function weights the payoffs of the neighbors with the ``friendliness'' $\rho_i$, i.e. the utility function $U_i$ of an agent $i$ is assumed to be $U_i = (1-\rho_i) P_i + \rho_i \langle P\rangle_i$, where $P_i$ is the payoff of individual $i$ and $\langle P \rangle_i$ the average payoff of the neighbors. For all agents, the initial friendliness is set to $\rho_i = 0$ (corresponding to a ``homo economicus'') and the initial behavior (``strategy'') is assumed to be defective (orange or red). The behaviors of all individuals are simultaneously updated based on a best response rule, which implies a behavior that maximizes the individual utility based on the behaviors of the neighbors. Only with a probability of 0.05 do individuals take other decisions than the best response rule suggests.

\par

To allow for evolution, while keeping population size constant, individuals die at random with the probability $\beta=0.05$, but are replaced with offspring of living individuals. The likelihood to give birth to offspring is proportional to the actual payoffs in the previous round (i.e. not proportional to the utility). With probability $\nu$, the offspring is born in the closest empty site to the parent, while with probability $(1-\nu)$, the offspring moves to a randomly selected empty site (here: $\nu = 0.95$). Offsprings inherit a trait called friendliness $\rho_i$ from their parents, which is subject to random mutations. (With probability 0.8 the offspring's friendliness is ``reset'' to a uniformly distributed random value between 0 and the friendliness $\rho_i$ of the parent, and with probability 0.2 it takes on a uniformly distributed value between $\rho_i$ and 1.) The local reproduction rate $\nu$ determines, whether a transition from a ``homo economicus'' with self-regarding preferences to a ``homo socialis'' with other-regarding preferences takes place, see Fig. 2 in \cite{Grund13}.} 

\label{FigCoopStrangers}

\end{figure}

Based on the above assumptions, the characteristic outcome of the evolutionary game-theoretical computer simulations is a self-regarding, payoff-maximizing ``homo economicus'', as expected. But this applies only to \emph{most} parameter combinations. When offspring tend to live close to their parents (i.e. intergenerational migration is low), a friendly ``homo socialis'' with other-regarding preferences results instead, see Fig. \ref{FigEmergence}.\footnote{The ``homo socialis'' may be defined as being sensitive to the social context, while the ``homo economicus'' is not. However, the other-regarding preference must not be reflected by a utility function that weights the payoffs of others, see Sec. \ref{trafficlight}.} This is quite surprising, since the above assumptions do not favor such an outcome. None of these rules promotes other-regarding preferences or cooperation in separation (i.e. they might be considered socially dysfunctional), but they are nevertheless creating socially favorable behavior in combination. This can only be explained as result of an interaction effect between the above rules. Another interesting finding is the evolution of ``cooperation between strangers'' (see Fig. \ref{FigCoopStrangers}), i.e. the emergence of the ``homo socialis'' does not require genetic favoritism. 

\par%

How can we understand this outcome? Mutations generate a certain level of friendliness by chance (``mutations''). This slight other-regarding preference (eventually giving the payoff of others an average weight of about 0.2) creates conditionally cooperative behavior, as postulated by Fehr and others \citep{Fehr99}. That is, if enough neighbors cooperate, a ``conditional cooperator'' will be cooperative as well, but not so if too many neighbours defect.\footnote{Conditional cooperators can exist for friendliness values $\rho_i$ with $(T-R)/(T-S) < \rho_i < (P-S)/(T-S)$ \citep{Grund13}, with the payoff values $P, R, S, T$ defined in the caption of Fig. \ref{FigCoopStrangers}. Below the lower threshold, an agent always defects and corresponds to a ``homo economicus''. Above the higher threshold, an agent is unconditionally cooperative (an ``idealist''). As a consequence, if all agents would have an identical friendliness $\rho$, one would expect a hysteretic system behavior as a function of the friendliness value. When the friendliness increases from low to high values, a discontinuous transition from defection to cooperation of everyone is expected when the value $(P-S)/(T-S)$ is exceeded, while a discontinuous transition from a cooperative to a defective behavior of everyone is expected when the friendliness value becomes smaller than $(T-R)/(T-S)$.} 
Unconditionally cooperative agents with a high level of friendliness are born very rarely, and only by chance. These ``idealistic'' individuals will usually be exploited, have very poor payoffs, and no offspring. However, if born into an environment where enough agents have a moderate friendliness and are conditionally cooperative, ``idealists'' with a high level of friendliness and unconditionally cooperative behavior can trigger cooperative behavior of the neighbors in a cascade-like manner. Under such conditions, high levels of friendliness are passed on to many offspring such that a ``homo socialis'' with other-regarding preferences emerges and spreads. This holds, because greater friendliness now tends to be profitable. On average, cooperators earn higher payoffs than defectors. Hence, the ``homo socialis'' can eventually outcompete the ``homo economicus'', while initially the ``homo economicus'' earns more (see Fig. \ref{FigEmergence}). Nevertheless, the friendliness levels are widely distributed, thereby explaining heterogeneous individual preferences, as empirically observed \citep{Murphy11}. In other words, everything from selfish to altruistic preferences exists.

In the situation studied above, where everyone starts as a defecting ``homo economicus'', no single individual can establish profitable cooperation, not even by optimizing decisions over an infinitely long time horizon. It takes a few ``friendly'' deviations to trigger cascade effects that eventually change the macro-level outcome. Therefore, it is important to recognize that a critical number of interacting individuals need to be friendly and cooperative by coincidence. The ``homo socialis'' would never evolve without the occurrence of random ``mistakes'' (here: the birth of ``idealists'' who are exploited by everyone). However, given suitable feedback effects, such ``errors'' enable better outcomes. Here, they eventually overcome the ``tragedy of the commons'' \citep{Hardin}, resulting in an ``upward spiral'' towards cooperation with high payoffs. Remarkably, this is not the outcome of an optimization process, but rather of an evolutionary process.

The simple evolutionary theory described in the paper discussed above contained very few fundamental and widely accepted assumptions, and might have the potential to form the basis of an integrated theoretical approach in the behavioral sciences. In accordance with a lot of work in social psychology, the ``homo socialis'' may be characterized by ``empathy'' in the sense that the ``homo socialis'' puts himself or herself into the shoes of others. More precisely, the ``homo socialis'' takes into account the perspective, interests, and success of others when taking own decisions.\footnote{However, taking account the interests of others must not necessarily be based on a weighted utility function, see the example on traffic light control in Sec. \ref{trafficlight}.} This effectively explains the fairness preference observed in behavioral economics \citep{Gueth82,Fehr99,Henrich01,Fehr03}. Note, however, that the ``homo socialis'' should not be imagined as a ``homo economicus'' sharing some payoff with others. The ``homo socialis'' decides differently from the ``homo economicus'', and that is why ``tragedies of the commons'' can be overcome.\footnote{One might say the ``homo socialis'' makes the individual and system optimum (more) compatible with each other.}

Decisions of the ``homo socialis'' are interdependent, in contrast to the ``homo economicus'', who takes decisions indendently without the consideration of effects on others. Therefore, the ``homo socialis'' may be characterized by the term ``networked minds''. This implies a complex dynamics of systems where many such decision-makers interact, with important economic implications \citep{HelbingRethinkingEconomics}. While methods from statistics should be good enough to characterize the ``homo economicus'', the description of the ``homo socialis'' requires methods drawn from complexity science.




\subsection{The FuturICT Initiative}

When non-linear interactions and network interdependencies prevail, economic systems are expected to be complex dynamical systems \citep{KrugmanSelfOrganizing1996,DayComplexEconomicDynamics1999,AuyangComplexSystemTheories1999,CopingComplexity2009,KirmanComplexEconomics,Brock97,Lux99, HelbingBaliettiChallenges,Keen13,Arthur2013ComplexityEconomics}. From ecology, climate research, and statistical physics it is known that such systems can not be well understood merely by analytical and econometric approaches, because of emergent system properties resulting from interactions between the system components \cite{Strogatz}.

A characteristic feature of complex dynamical systems is that they may not necessarily be in equilibrium \citep{EmergentMacroeconomics,HelbingWittetal}, and that a representative agent approach may not work as well \citep{BeyondtheRepresentativeAgent,Deguchi2004,HelbingSSO}, partly because of strong correlations and cascade effects \citep{BattistonLiaisonsDangereuses2012, BankruptcyCascades,HelbingGloballyNetworkedRisks}. Therefore, the FuturICT initiative (see \url{http://www.futurict.eu}) has recently proposed to combine socio-economic approaches with complexity science, massive computer power and Big Data to analyze, model and simulate techno-socio-economic-environmental systems. The initiative aims to develop a better understanding of the highly interdependent and densely networked world we are living in, to increase its sustainability and resilience.

The FuturICT initiative points out that the globalization and the increasing connectedness in human-influenced (``anthropogenic'') systems has increased the complexity of our world and also created systemic risks \citep{HelbingGloballyNetworkedRisks}. It is argued that, due to emergent phenomena, densely connected and strongly interacting systems cannot be understood from the properties of the components and the behaviors of separately deciding agents \citep{HelbingFuturICTParticipatory}. This implies the need to fundamentally change our conceptional approach, namely to move from a component-oriented perspective to an interaction-oriented, systemic perspective. Consequently, policy implications may look quite different from what the theory of independent decision-makers suggests, but this may help to overcome some long-standing problems \citep{SalzanoColander,HelbingRethinkingEconomics,HelbingSSO}.

In order to ensure that our knowledge can keep up with the speed of change of our complex, globalized world, the FuturICT initiative also proposes to establish a ``Global Systems Science'' aiming at a ``grand integration of knowledge'' by combining the best knowledge of the social, natural and engineering sciences in a large-scale and truly interdisciplinary effort \citep{HelbingGloballyNetworkedRisks}. Despite competition, I believe that scientific curiosity can promote a respectful interaction and cooperation that can generate insights on all sides. For this, one should promote a culture of cross-disciplinary appreciation.




\section{Differences between the ``Homo Socialis'' and the ``Homo Economicus''} \label{Differences}

One may argue that the ``homo socialis'', who considers the payoff of others, has been covered by rational choice theory already \citep{Bergstrom1999,InterdependentUtilities2001}, since it has been increasingly recognized that (many) people do not just optimize the own payoffs (as shown, for example, by Ultimatum Game experiments, see \cite{Gueth82, Henrich01}). To take this into account, it has been assumed that individual preferences are part of the utility function and that altruistically behaving individuals enjoy helping others, while the others do not. However, when the utility function is treated as individual rather than universal quantity, a rational choice theory based on utility maximization loses much of its predictability and strength. Then, the utility function cannot be calculated from first principles, but it must be statistically fitted from data separately for each individual. In contrast, the above theory of the ``homo socialis'' provides an evolutionary explanation of the individual utility function, which is a major advantage.

In fact, both, the ``homo economicus'' and the ``homo socialis'' assume self-determined individual decisions, just according to different utility functions. At first glance, it may appear that this difference is not important. However, when the ``friendliness'' parameter in the utility function is changed continuously, one eventually crosses a ``tipping point'' beyond which the resulting individual and system behavior look dramatically different. In other words, when the friendliness parameter is varied, the system behavior changes discontinuously. Individual optimization attempts will only simultaneously create individual prosperity and maximum social benefits beyond a certain level of ``friendliness'', i.e. for the ``homo socialis''.

In other words, in social dilemma situations, Adam Smith's principle of the ``invisible hand'' works only for the ``homo socialis'', but not for the ``homo economicus''. Without suitable institutional settings or regulations, the ``homo economicus'' will run into a ``tragedy of the commons'' with very poor payoffs for the great majority. Note that the principle of the ``invisible hand'' would work for the ``homo economicus'' as well, if the underlying assumptions of the First Theorem of Welfare Economics were fulfilled, i.e. if all individuals would interact in one shared market without transaction costs and externalities \citep{Arrow54,DebreuTheoryOfValue}. However, these assumptions are quite restrictive, and the macro-level outcome depends on them in a sensitive manner. For example, non-centralized network interactions and transaction costs can manifestly change the outcome \citep{Coase1960,RocaHelbingPercolationChapterInSocialSelfOrganization}.

As the ``homo economicus'' decides strictly according to self-interest, the coordination of decision-makers and efforts to overcome ``tragedies of the commons'' or market failures requires some kind of top-down regulation. This results in a steadily growing number of laws and regulations and costly investments into regulatory and sanctioning institutions (such as police, courts, market regulators, etc.). This can significantly decrease systemic efficiency. In addition, regulation tends to reduce diversity, which can affect innovation in a negative way \citep{page,Inno,Inno2,Inno3,Helbing05b}.

In comparison with the ``homo economicus'', the ``homo socialis'' is able to cope with social dilemma situations based on local interactions. I will characterize this as bottom-up \emph{self-regulation}, considering that the emergent level of ``friendliness'' overcoming the ``tragedy of the commons'' results from a decentralized, evolutionary process. Since top-down regulation (as required for the ``homo economicus'') is just the opposite of bottom-up self-regulation (as typical for the ``homo socialis''), it is misleading to say that the ``homo socialis'' is just a variant of the ``homo economicus'' or vice versa. Speaking metaphorically, the ``homo economicus'' and the ``homo socialis'' must be distinguished like night and day. This is mainly due to the discontinuous transition in the systemic outcome when the ``friendliness'' increases beyond a certain critical level (``tipping point'') (see footnote 16). 
Passing the tipping point is like switching the light on or off, with dramatic impacts on the average payoff (see Fig. \ref{FigEmergence}).




\section{Socionomics: A Promising New Field of Research} \label{Socionomics}

The existence of two different kinds of people, the ``homo economicus'' and the ``homo socialis'', has relevant practical implications: A second body of work must be developed for the ``homo socialis'' to supplement the extant economic body of work for the ``homo economicus''. I will call the science of the ``homo economicus'' - ``economics'', and the science of the ``homo socialis'' ``socionomics'' or also ``economics 2.0'' in order to reflect that the Web 2.0 is a key factor promoting a ``participatory market society'' \citep{oreilly2007web}.  It is important to note that our understanding of socionomics does not assume emotionally or irrationally acting agents. In the simulations underlying Figs. \ref{FigEmergence} and \ref{FigCoopStrangers}, the ``homo socialis'' decides rationally, trying to optimize the individual utility, but this utility also reflects the externalities of others. In social dilemma situations, the ``homo socialis'' can outperform the ``homo economicus'' through reaching higher payoffs.




\begin{figure}[htbp]

	\centering

    \includegraphics[width=0.6\textwidth]{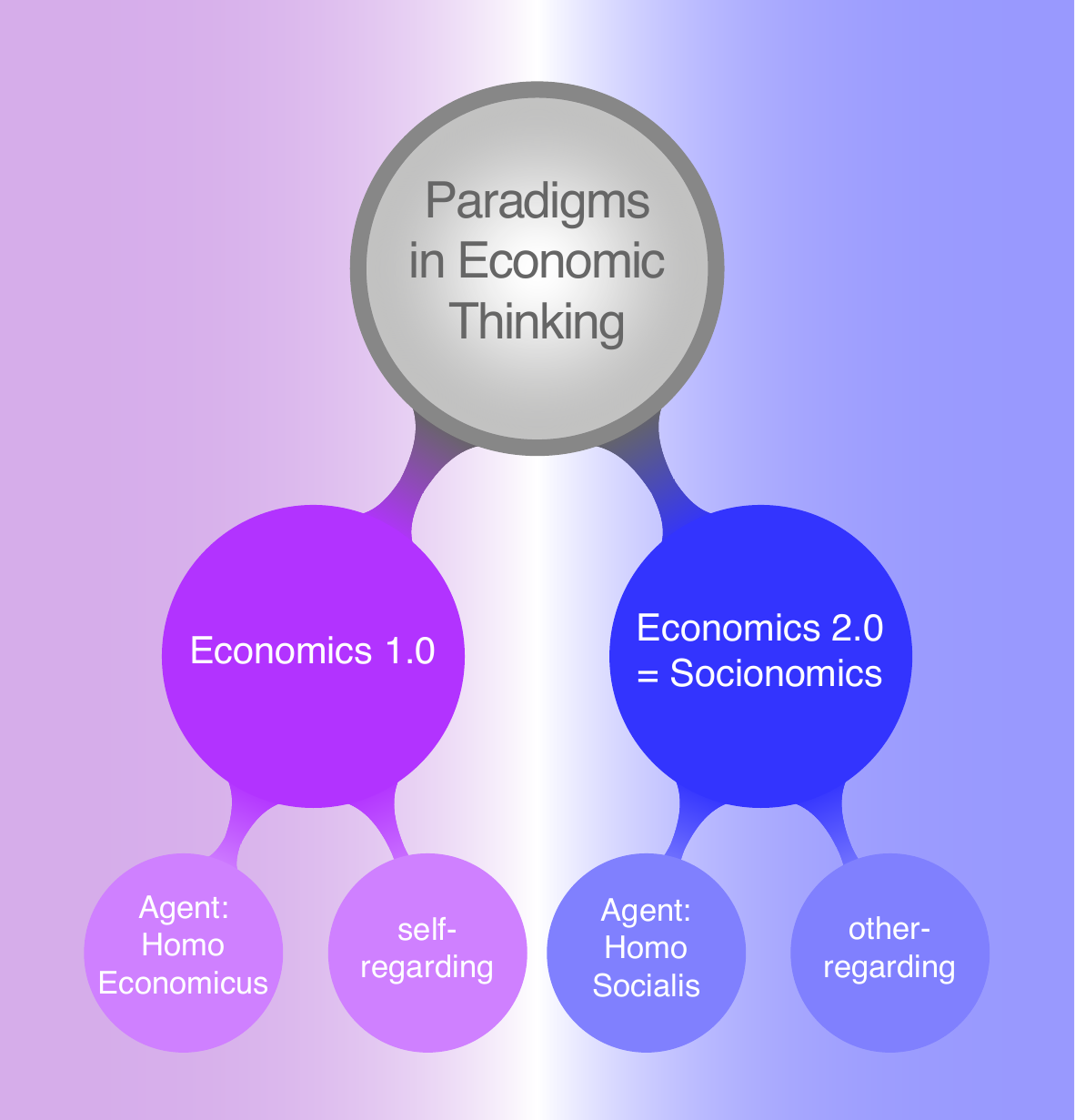}

    \caption[]{\textbf{Illustration of the two distinct economic paradigms.} The difference between the ``homo economicus'' and the ``homo socialis'' is that the latter takes into account the interests of others when making decisions, which implies interdependent decisions or \emph{``networked minds''}. The different nature of the ``homo socialis'' leads to a complex dynamics and another macroscopic outcome than expected for the ``homo economicus''.

In the case of public goods problems, for example, interactions of agents with strictly self-regarding preferences will lead to \emph{``tragedies of the commons''}, while the self-regulation of the ``homo socialis'' can overcome this undesirable state and foster cooperation, leading to higher individual and social benefits.

Due to the different system dynamics and different systemic outcomes, both types of agents cannot be described by the same body of theory. They require separate sets of literature and different institutions.}

    \label{Paradigms}

\end{figure}

\subsection*{Socionomics, and What Distinguishes it from Economics}

In contrast to prevailing work in sociology, socionomics includes the study of suitable institutions and operational principles for future market and exchange systems.\footnote{I would like to point out that socionomics has certain points in common with other emerging research directions such as \emph{evolutionary economics}, \emph{network economics},  \emph{complexity economics}, \emph{behavioral economics}, or \emph{socio-economics}, to mention just a few.} Market and exchange systems fitting the ``homo economicus'' and the ``homo socialis'' will be different. In the same way as economic market systems are called ``economies'', I will call market systems based on socionomic principles ``socionomies'' or ``economies 2.0''. I will also use the term "participatory market societies'', as socionomies bring economic activities together with a social orientation. The participatory character of socionomies will become clearer when I discuss the role of ``prosumers'', i.e. of co-producing consumers in Sec. \ref{ProsumersEntrepreneurs}.




\begin{figure}

	\centering

    \includegraphics[width=0.8\textwidth]{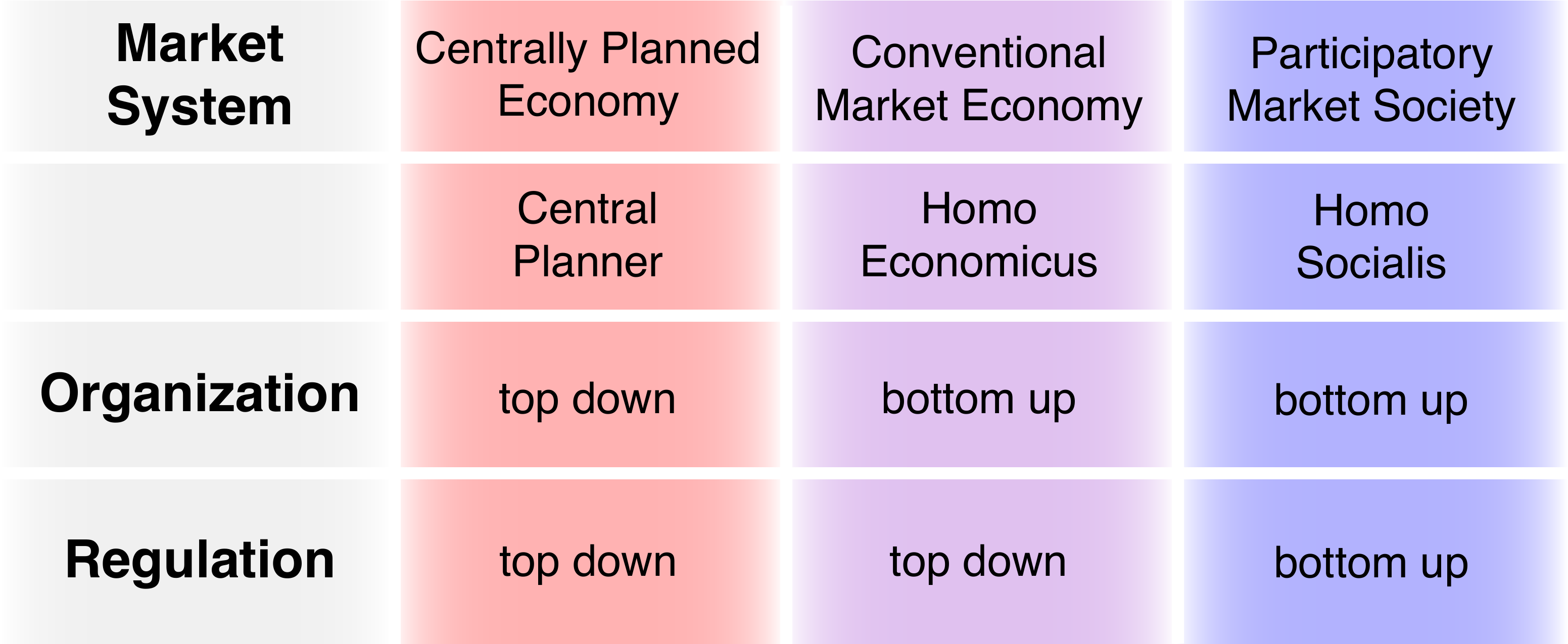}

    \caption[]{\textbf{Comparison of different ways of organizing and regulating markets, based on a (benevolent) ``central planner'', the ``homo economicus'' or the ``homo socialis''.} The organization and regulation of the ``homo socialis'' needs to be distinguished from both, centrally planned an conventional market economies. Hence, I propose to devote a new branch of economics called \emph{``socionomics''} or \emph{``economics 2.0''} to the study of the system dynamics, outcomes, and suitable institutions of the ``homo socialis''. The related socio-economic systems are ``participatory market societies''. These systems are now emerging and spreading due to the Web2.0, particularly thanks to social media platforms.}

    \label{EconomicConcepts}

\end{figure}

But what might be distinguishing features of socionomies as compared to economies?




\begin{itemize}

\item In socionomies, self-regarding optimization is replaced by individual decisions that take into account the implications for others (``externalities''), thereby supporting mutual coordination and cooperation. 

\item Therefore, in socionomies, agents would have cooperative motives in addition to competitive ones. This is sometimes characterized by the term ``coopetition''\footnote{see http://en.wikipedia.org/wiki/Coopetition}.

\item In socionomies, regulation can happen in a bottom-up way, which is called ``self-regulation''.

\item In socionomies, social dilemma situations can also be overcome without financial incentives and fines, for example, by reputation systems (see Sec. \ref{ReputationSystems}).

\item In socionomies, sharing information and other resources pays off, as this supports coordination and cooperation.

\end{itemize}





\begin{figure}

	\centering

    \includegraphics[width=1.0\textwidth]{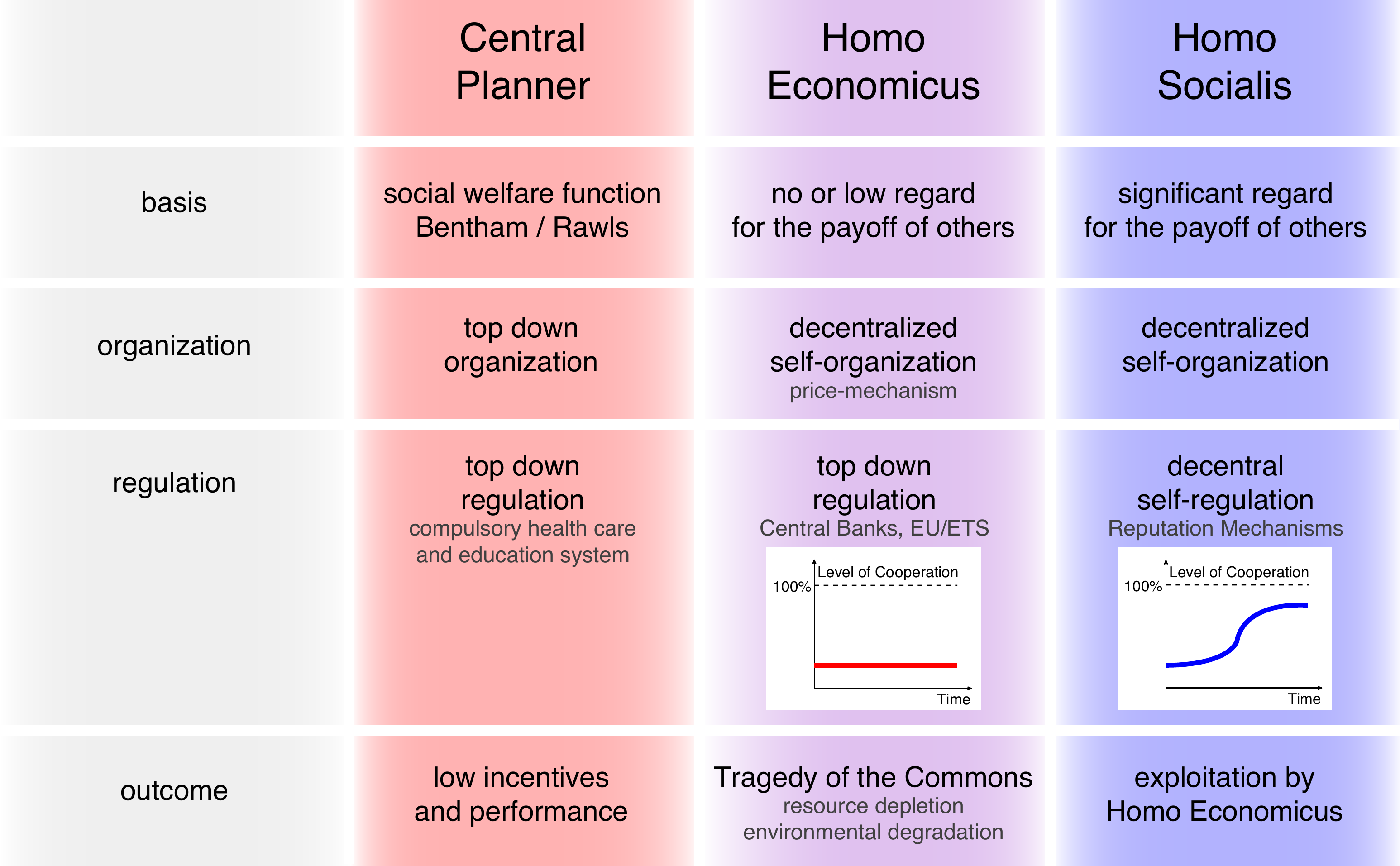}

    \caption[]{\textbf{Table highlighting the differences between the nature of a central planner, the ``homo economicus'', and the ``homo socialis'', including their different system dynamics and systemic outcomes.} The First Theorem of Welfare Economics\citep{Arrow54,DebreuTheoryOfValue}, which states that competitive markets lead to an efficient allocation of resources, only holds if there is a complete set of markets and all market participants are price-takers, and there are neither externalities nor transaction costs.

As conventional market economies tend to create negative externalities, top-down regulation by a strong state is needed to counter related market failures. Examples of negative externalities are resource depletion and environmental pollution, which eventually lead to the creation of regulatory institutions such as the emission trading scheme or financial regulatory authorities. As regulation tends to be costly for regulators and regulated agents, bottom-up self-regulation is expected to unleash large creative potentials of ``networked minds'', which are currently not fully used (see Sec. \ref{ParticipatoryMarketSociety}). The performance gains may well be comparable to the gains when moving from a top-down regulation to a conventional market economy building on self-organization (see Sec. \ref{trafficlight}).

``Socionomies'' (also called ``economies 2.0'' or ``participatory market societies''), are characterized by high levels of cooperation, established through self-regulation. To protect a ``homo socialis'' from being exploited, it is important to know whether one is interacting with another ``homo socialis'' or a ``homo economicus''. Therefore, suitable institutions such as reputation mechanisms need to be established, which allow like-minded agents to interact successfully, and differently minded agents to cope with each other. Examples for socionomic systems are Wikipedia, Open Streetmap, open source software communities (such as Linux or GitHub), and social media platforms.

Note that the ``homo socialis'' should not be confused with someone, who is redistributing gains. A redistribution strategy (via donations, or taxes and social benefit systems) might be able to ``heal'' or reduce problems such as an ``unhealthy'' degree of inequality, but it would usually not be able to avoid or overcome ``tragedies of the commons''. Hence, redistribution is inefficient and often not a suitable solution for social dilemma situations (see also footnote 27). The main difference to the ``homo economicus'' is that the ``homo socialis'' decides more ``responsibly'' or ``wisely'', considering not only the own interests, but also those of others when taking self-determined decisions. If enough people do this, other-regarding preferences pays off.}

    \label{FigDifferences}

\end{figure}

As socionomics is emergent, the above points need to be investigated with simulation models and experiments to identify the power and limitations of the new paradigm of the ``homo socialis'' and the institutions needed (see Sec. \ref{InstitutionsHomoSocialis}).




\subsection{What One Can Learn from Traffic Light Control}\label{trafficlight}




\begin{figure}

    \centerline{\includegraphics[width=0.8\textwidth]{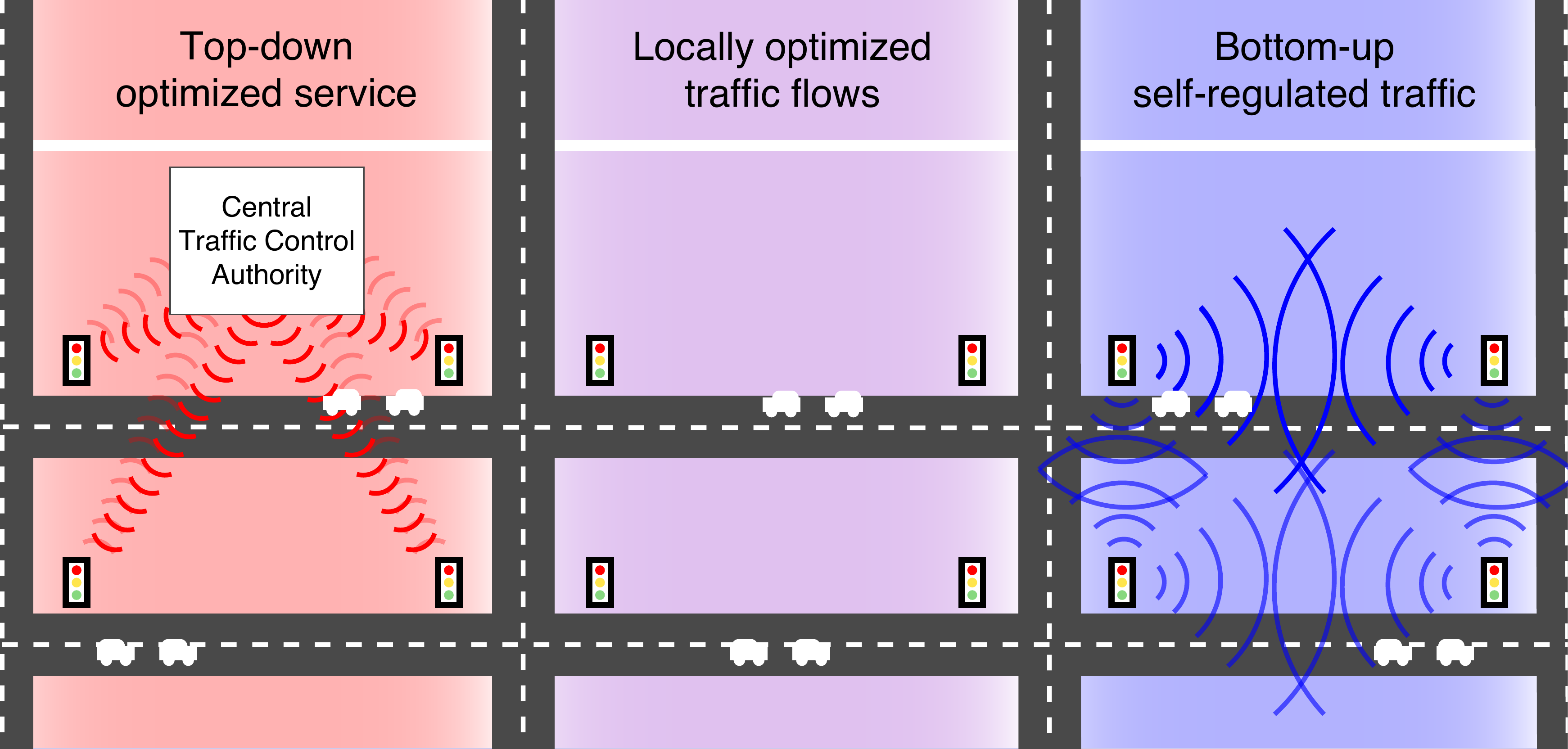}}

    \caption[]{\textbf{Schematic illustration of three kinds of traffic light control:} a ``central regulator'', a strategy minimizing local waiting times (``homo economicus'' approach) and a strategy considering externalities on neighboring intersections (``homo socials'' approach).}

    \label{FigTrafficLights}

\end{figure}

Like the preferences of economic agents, urban traffic flows are characterized by externalities (as they negatively interfere with each other). Therefore, when discussing different ways of organizing and regulating markets, one can learn much from urban traffic light control. For this purpose, let us treat road intersections as agents that are networked with each other. Then, if we let all intersections optimize the local traffic flows independently of each other, this corresponds to the approach of a ``homo economicus''. If the intersections take into account the traffic flows at neighboring intersections in their control decisions, this corresponds to the ``homo socialis''. I will also compare this with the approach of a central regulator, who tries to optimize traffic flows in the entire city in a top-down way.

The typical goal of traffic control is to minimize congestion,  environmental impact, and travel times. Today, decisions are usually taken top down by a traffic control center, corresponding to a \emph{central regulator}. However, typically, the mobility demand changes more quickly than the control strategy can be adjusted, with the consequence that the optimization of traffic flows often leads to suboptimal results.

In the following paragraphs, I will explain how and why a \emph{self-regulation} approach based on a flexible, adaptive response to local traffic flows can be more successful than the top-down control by a central regulatory authority. To control traffic lights, many cities today use supercomputers trying to identify the optimal solution for the system and to implement it like a ``benevolent dictator''. A typical solution  creates ``green waves'' by synchronizing neighboring traffic lights. However, in large cities, even supercomputers are unable to manage strict traffic light optimization in real-time due to the algorithmic complexity of the optimization problem (``NP completeness''). The number of parameters and the possible combinations of their values are so large that the required computational time ``explodes'' with the number of traffic lights.

Instead, traffic light control schemes are usually optimized offline for representative (average) traffic flows at a certain weekday and time, or for certain events (e.g. football games)\footnote{This is also the approach that the mean value approximation behind the representative agent approach would suggest.}. The corresponding control schemes are then activated in the situations for which they were optimized. In addition, these may be adapted to the actual traffic situation by extending or shortening green phases. However, at a particular intersection the periodicity (i.e. the order of green phases serving the incoming roads) is usually kept the same during the operation of a control scheme. While the adaptation of traffic phases improves the traffic light control, it must be recognized that the concept of ``representative'' traffic flows is very misleading: the variability of the number of vehicles waiting at a traffic light is about as large as the average value. This implies a sub-optimal performance of the central regulator approach.

\begin{figure}

\centerline{\includegraphics[width=0.85\textwidth]{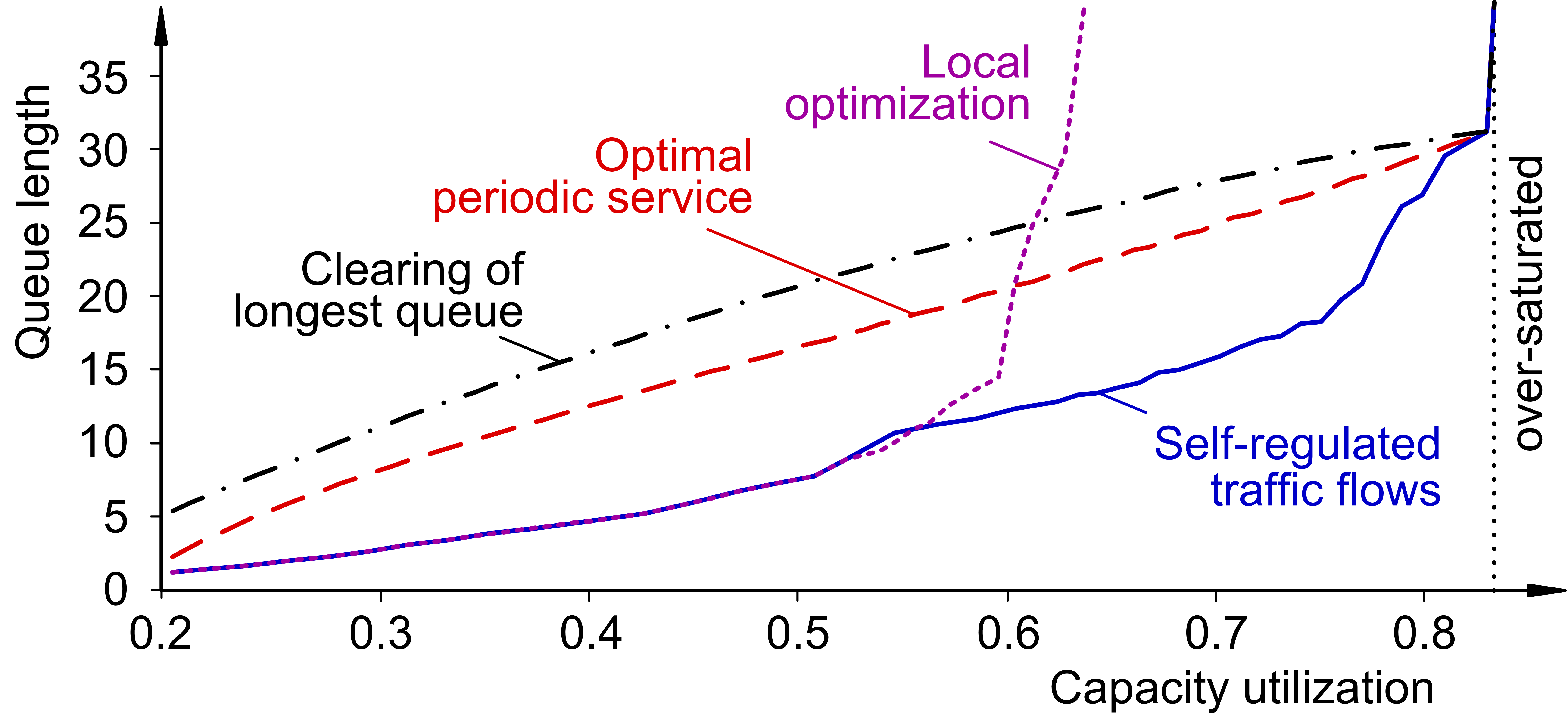}}

\caption[]{\textbf{Comparison of the performance of a central regulator, local optimization of traffic flows as a ``homo economicus'' would do, and other-regarding self-regulation, considering the impact on neighboring intersections.} 

The graphics shows average queue lengths resulting for these different traffic control approaches (after \cite{LaemmerSFI}). Always clearing the longest queues performs worse than a periodic service that is optimized in a top-down way, as a central regulator would do. A local minimization of waiting times, corresponding to a situation where each intersection acts like a ``homo economicus'', outperforms a periodic service at low utilizations of the theoretical intersection capacity, but it creates long queues and spill-over effects at high utilizations. A decentralized self-regulation with other-regarding decision rules, corresponding to the approach of a ``homo socialis'', performs best. It also minimizes waiting times locally, but this principle is over-ruled by the clearing of critical queues, which might block neighboring intersections by spill-over effects. The self-regulation approach also outperforms an optimized cyclical service, where the green times are adapted to the respective traffic situation \citep{LaemmerSFI}.

}

\label{TrafficLaemmer}

\end{figure}

Let us now compare the central traffic light control with the ``homo economicus'' strategy, where each intersection strictly optimizes local traffic flows independently by minimizing average waiting times \citep{LaemmerHelbingJSTAT}. We assume that each intersection measures not only the outflows of the incoming road sections, but also receives information about the inflows into these road sections that originate at the neighboring intersections. This information exchange between neighboring intersections allows short-term predictions of the arrival times of vehicles. Hence, the ``homo economicus'' strategy can respond to this prediction in a way that tries to keep vehicles moving, thereby minimising waiting times. This self-organizes the local traffic flows, but the principle of the ``invisible hand'' works only up to a certain traffic volume \citep{LaemmerHelbingJSTAT}. Long before the intersection capacity is reached, the average queue length tends to get out of control, as some road sections with small traffic flows are not served frequently enough (see Fig. \ref{TrafficLaemmer}).\footnote{In other words: minimum waiting times are generally not compatible with minimum queue lengths and vice versa.}  This will create spillover effects and obstructions of upstream traffic flows, such that congestion quickly spreads over large parts of the city in a cascade-like manner. The resulting state may be compared with a ``tragedy of the commons'', as the available intersection capacities is not efficiently used.




\par%

In order to outperform the central regulator over the whole range of traffic demands (utilizations) an intersection can handle, one must extend the ``homo economicus''-based self-control approach to a self-regulatory approach. Specifically, the rule of waiting time minimization must be combined with a second rule, according to which a vehicle queue must be cleared immediately, whenever it reaches a critical length (a certain percentage of the road section).  This second rule avoids spill-over effects that would obstruct neighboring intersections and therefore acts in an other-regarding way, as a ``homo socialis'' would do. The self-regulation strategy uses occurring gaps as opportunities to serve other traffic flows. So, rather than planning green waves by synchronizing traffic lights in a certain rhythm, self-regulated traffic lights take into account the traffic situation at neighboring intersections, just as a ``homo socialis'' would do. This enables the coordination of neighboring traffic lights which can spread over many intersections.

In summary, the self-regulation approach coordinates traffic lights and traffic flows based on local interactions. While the self-regulation is based on a decentralized bottom-up approach rather than a centralized top-down approach, it surprisingly reaches a superior performance \citep{LaemmerSFI}. Studies comparing self-regulated traffic lights with state-of-the-art traffic control approaches show that smaller waiting times can be simultaneously reached for individual traffic, public transport, pedestrians, and bikes \citep{LaemmerSFI}. Self-controlled traffic also reduces environmental pollution without having to impose travel restrictions, and it also makes travel times more predictable.

From the above example, we can draw a number of important conclusions:




\begin{enumerate}

\item In complex systems with strongly variable and largely unpredictable dynamics, bottom-up self-regulation can outperform top-down optimization by a central regulator (an urban traffic control center). This also holds when input data are comprehensive and reliable.

\item Strict local optimization according to the principles of a ``homo economicus'' may create a highly peforming system within a certain parameter range, but tends to fail when interaction effects by traffic flows of neighboring intersections are strong and the optimization at each intersection is performed separately, in a self-regarding way.

\item When taking into account the situation of interaction partners (``neighbors''), as an other-regarding ``homo socialis'' would do, a high system performance can be reached even for strong interaction effects because of coordination between neighbors.

\end{enumerate}


In conclusion, a central regulator fails to be efficient for lack of computational capacity, which is required to handle the algorithmic complexity of a NP-hard optimization problem within  large system in real-time. Applying the principle of the ``homo economicus'' on a local level, fails due to lack of coordination. In contrast, the approach of the other-regarding ``homo socialis'' can overcome both problems by combining self-organization with self-regulation. A self-regulation based on suitable local interactions can produce resource-efficient solutions delivering high system performance.\footnote{The recent trend towards replacing many signalized intersections by roundabouts and changing urban spaces regulated by many traffic signs in favor of designs supporting a considerate interaction and self-organization of different traffic participants suggests an on-going paradigm shift from regulation to self-regulation \cite{sharedspace}.} These solutions also tend to be resilient to perturbations (such as accidents, road works, etc.). Again, the ``homo socialis'' can make the principle of the ``invisible hand'' work under conditions, where it fails for the independent optimization attempts of a ``homo economicus'' (see Fig. \ref{TrafficLaemmer}).

Many of the above conclusions also seem to be relevant for socio-economic systems, when agents have incompatible interests that cannot be satisfied at the same time.


\section{Institutions for the ``Homo Socialis''} \label{InstitutionsHomoSocialis}


\subsection{Mechanisms to Overcome Social Dilemma Situations}

Institutions for market systems should not just work under ideal (``good weather'') conditions, but also under difficult ones. They should be suited not only for best-case scenarios, but also for worst-case scenarios, in as much as this is possible. Therefore, in the spirit of the self-regulation approach, we are looking for:

\begin{enumerate}

\item rules to handle externalities of decisions, protecting others from exploitation and compensating them for damage,

\item frameworks to coordinate between individuals and projects with different preferences and interests.

\end{enumerate}

In particular, we are searching for institutions that work in the case of social dilemma situations. From game theory, it is known that cooperation is endangered by interaction with defectors \citep{FehrAltruisticPunishment}. \\

As the ``homo economicus'' is always expected to defect when deciding according to a best response rule (maximizing the payoff given the current behaviors of the neighbors), cooperation by the ``homo socialis'' must be somehow protected from exploitation by the ``homo economicus''.  In other words, the ``homo socialis'' is not expected to thrive  in institutional settings designed for the ``homo economicus'' such as homogeneous global markets, which destabilize cooperation in social dilemma situations \citep{HelbingRauhutetalJournalMathematicalSociology,HelbingGloballyNetworkedRisks}. I would like to stress again that any situation, in which a market participant can be cheated, establishes a social dilemma situation. Nevertheless, there are solutions to the problem.

Evolutionary game theory has identified a number of biological and social mechanisms that can create suitable institutional settings for cooperation in social dilemma situations. These includes genetic favoritism and direct reciprocity (``I help you, if you help me'') \citep{NowakEvolutionary}, but also punitive institutions (such as the criminal justice system, and legal and regulatory authorities). However, all these mechanisms have undesired side effects. Genetic favoritism impedes an open global exchange and leads to ethnic conflict. Direct reciprocity is hardly compatible with global openness to new trading partners, and could also incentivise corruption. Punishment institutions are costly and create inefficiencies, as discussed in Sec. \ref{Differences}. In Sec. \ref{homosocialis}, cooperation spreads due to a naturally resulting spatial segregation between cooperators and defectors, when intergenerational migration is low \citep{Grund13}. Another mechanism is based on reputation systems \citep{ross,Milinski02}, which transfers the functional principle of social communities to a globalized world. Such reputation systems work also in a decentralized way \citep{KamvarEigentrust2003, BucheggerReputation2004, XiongPeertrust2004,Zhou,Mundinger}.

\subsection{User-Centric, Multi-Criteria Reputation Systems and Incentives} \label{ReputationSystems}

The purpose of the reputation system is to avoid costly transaction failures arising from mismatched expectations regarding the kind of economic exchange \citep{WinterRauhutHelbing}. In other words, reputation systems can help different types of agents such as the ``homo socialis'' and the ``homo economicus'' to cope with each other. Reputation systems are currently spreading on the Web. A typical example is ebay. Reputation and recommender systems are now used to evaluate goods, service providers  and news. It has been shown that ebay sellers with a higher reputation can sell products at a higher price, i.e. a good reputation pays off \citep{Przepiorka2013ReputationEbay}. This incentivises reputation-enhancing behaviour. In addition to price-based competition, recommender and reputation systems also endorse quality-based competition, i.e. features such as responsible environmental and social production conditions become more relevant and rewarding.

An important question to ask is: who would decide the reputation of someone or something? To successfully cope with a system of high complexity (see Sec. \ref{trafficlight}), it seems necessary to build on a bottom-up self-regulation approach. Therefore, the reputation should be decided decentrally, by each decision-maker, based on his or her own preferences and values. In other words, it should not be a company or another institution that determines the reputation, but the individuals.\footnote{The problems of recommender systems, which cannot be configured by the users themselves, have been discussed in the book ``Filter Bubble'' \citep{Pariser}. One of them is the manipulation of decision-makers, because the ``wisdom of crowds'' can be seriously undermined by feeding back opinions of others \citep{Lorenz11}. This can reduce socio-diversity as well \citep{GoogleAsGod} and, thereby, affect the functionality of fundamental institutions that societies are built on. Democratic and market institutions require independent decision-making to function well.} We should also consider that criteria determining high quality for some, may represent poor quality for others. Thus ideally, recommender systems should be multi-dimensional, based on a variety of criteria. Decision-makers can then define filters determining reputation based on their own criteria.\footnote{As a consequence, someone may have a good reputation in some areas and no or bad reputation in others, depending on the values and viewpoints of the potential interaction partners. Therefore, a user-centric, multi-criteria reputation system can help one to find like-minded people. Note that reputation data do not have to be public, but may also be provided on request.} Moreover, people could share their own reputation filters with others, who may modify them according to their own perspective. Therefore, opening up recommender data to everyone would enable the evolution of an ``information ecosystem'', in which customized information filters would steadily improve over time. A similar reputation principle could also be applied to money.




\subsection{Qualified Money\label{qualifiedMoney}}

One interesting -- and somewhat speculative -- question is what would happen, if we applied the principles of reputation systems to money itself \citep{Moreton2003TradingInTrust}, i.e. if each unit of money could earn a reputation, depending on its origin and transaction history.\footnote{I would like to underline that this section has currently a preliminary and speculative character, as it first needs to be grounded on computational and empirical evidence before it can be practically applied. However, the exploration of complementary, alternative, and backup financial systems seems to be in place.}  Then, units of money could be treated as separate stocks. Thus money would be related not just with a quantity, but also with some qualities. This would make money multi-dimensional, akin to  feedback and exchange systems in biological and ecological systems, or also social systems \citep{FiskeStructures}. I call this concept ``qualified money''.

Note that the one-dimensionality of today's money has a number of problematic features. It basically implies that there can only be ups and downs. The bubbles and crashes that financial systems experience over the last  hundreds of years \citep{GraeberDebt,GlynDaviesHistoryOfMoney} may be a consequence of this. For a complex dynamical system such as the financial system to work well, it is important that there are enough parameters to allow the system to adapt.\footnote{Note that trying to reach many different, non-aligned goals with just one control variable is an ``ill-defined'' (unfeasible) control problem. Assume that one tries to encourage $k$ desirable behaviors and discourage $l$ undesirable ones with corresponding positive and negative tax incentives, then the overall effect will be rather unspecific and most of the goals are unlikely to be reached.} 
Currently, the financial system suffers from a lack of such parameters.

If we had ``qualified money'', a conversion factor would apply, which would determine the value of money together with its quantity. The conversion factor would depend on the qualities of the respective money units, which would be given by multiple reputation values. Hence, the conversion factors establish adaptive parameters. Therefore, Euros in a certain country could gain a higher or lower reputation (and value) than in other countries. If a country would suffer from an economic depression, the conversion factor would decrease, and the corresponding devaluation of money would help the country to solve its problems by inflation. In other words, the international financial system would have enough degrees of freedom for self-regulation to work.

As indicated before, the qualifiers could also be made dependent on the transaction history. If money were history-dependent, then money generated in certain ways (e.g. by environmental-friendly production) would gain additional value, if a customer cares about it. This would create incentives to invest into quality, not just quantity. Therefore, the downward spiral leading to ``tragedies of the commons'' could be overcome. There would be a competition for money of higher quality. Note that the properties entering the qualifiers (and those that are not made transparent or evaluated) would be decided in a bottom-up way, first of all by the confidentiality settings of the owner and second by letting potential customers choose their own quality filters, as described for the multi-dimensional reputation system before. This would establish an implicit negotiation process between sellers and buyers, i.e. it would give customers more influence on price formation than just by a decision to buy a product or abstain from it. In this way, customers would become a ``voice'' allowing them to inform producers about the qualities they value and the factors they care less about. If this would influence the corresponding conversion factors strongly enough, it would be an effective way of improving quality standards. In a such a way, ``qualified money'' could also help to increase sustainability.

\subsection{Platforms for Participation, Exchange and Mutual Compensation}

Another important institution for the ``homo socialis'' are participatory platforms that support a trusted and fair  exchange \citep{Montjoye}, as required to coordinate with each other and to consider the interests and preferences of others. Such platforms should provide tools for cooperation, allowing people to set up joint projects, communicate and collaborate. Ideally, the platform would support everything from the scheduling of processes to supply chain management and accounting. The functionality should be easy to use at an affordable price with open and interoperable interfaces, while providing secure information storage and exchange for sensitive data. New information and communication technology (ICT) and the data they produce and manage will also make it possible to determine and quantify harmful interactions, and to reduce them by compensation mechanisms (such as appropriate payments). This possibility arises, in particular, as the ``Internet of Things'' spreads. 

The FuturICT initiative has recently proposed to build a ``Global Participatory Platform'', which would eventually allow everyone to set up collaboration projects with others, underpinned by Open Data \citep{ShumParticipatoryPlatform12}. Such a platform could also be used for eGovernance applications such as the payment of taxes, health and other insurances. From the perspective of self-regulation, it seems to make sense that the people involved in participatory decision-making would be those who are significantly affected by the respective externalities and those who care about the subject. In the future, this principle might be applied to economic and political decision-making alike.

\subsection{Innovation Accelerator}

To promote innovation, it is important to have incentives to invest efforts into innovation. Today, this is done by protecting intellectual property by patents. It often appears, however, that patents can also obstruct innovation. It would therefore be desirable to find a mechanism to reward innovations automatically, whenever someone's idea is used.\footnote{In this connection, Altmetrics (\url{http://www.altmetric.com/}) measuring page views and other web indicators might play an important role in the future.} If we had a way to measure the value of micro-innovations, we could innovate in a much more collaborative and open way. Whoever would have a good idea could contribute to a difficult invention and get a reward for it. Building  on crowd sourcing and swarm intelligence in such a way could largely speed up innovation, and also lead to higher-quality results \citep{InnovationAcceleratorVanHarmelen2012}. Rewards could either be micropayments per use (as spotify does it, for example) or non-monentary incentives (such as citations in science or reputation points), or both.

\subsection{``Socionomics'', or ``Economics 2.0''}

Last but not least, one important institution to promote the ``homo socialis'' would be a scientific discipline to study the implications of this concept and suitable institutional settings \citep{Schotter,SchubertWangenheim}. As pointed out before, this calls for a field such as ``socionomics'' or ``economics 2.0'' \citep{Eco2}. Note that one particular challenge will be to study the principles of successful self-regulation within complex systems. This will have to go beyond mechanism design \citep{HurwiczReiter}, taking into account evolutionary principles \citep{GeneticAlgorithms} and thereby endogenizing innovation into the system dynamics. Inspiration could also be drawn from ecological \citep{HaldaneMaySystemicRisk2011}, immune \citep{floreano}, or social systems \citep{HelbingSSO}.

Evolutionary principles would be able to identify improved rules for self-regulation in a truly bottom-up way. The case of traffic light control (see Sec. \ref{trafficlight}) has shown that small changes of rules may fundamentally change the systemic outcome and performance (such as large-scale congestion as compared to free traffic flows). When a system is sensitive to details, self-regulation must include rules that stabilize the application of successful rules (e.g. avoid a self-regarding modification of rules). However, over-rigidity would affect innovation and, therefore, the ability to find even better sets of rules. Evolutionary systems are therefore characterized by checking out diverse variants (thanks to innovation-generating ``mutations'') in combination with autocatalytic or reproduction mechanisms ensuring that better variants spread more quickly. As a consequence, a decentralized approach is an important precondition for success, while the implementation of homogeneous rules in large areas is expected to reduce diversity and innovation.


\section{Economies 2.0 as Participatory Market Societies\label{ParticipatoryMarketSociety}}


In the previous sections, I have discussed some theoretical arguments supporting a new organization of market exchange. In this section, we will see that markets have already started to re-organize in the suggested direction. For example, reputation and recommender systems are now found all over the Web 2.0. Likewise, Social Media are  providing participatory platforms, which are increasingly used to set up collaborative projects. The example of facebook demonstrates that even with free platforms  one can generate considerable economic value. The next paragraphs give a better idea, how future information and communication technologies are expected to change the lives of workers, consumers and entrepreneurs.




\subsection{Prosumers and the Future Role of Entrepreneurs\label{ProsumersEntrepreneurs}}

Today, we are still living in a world of slowly adapting institutions, which (in the best case) are trying to take optimal decisions for many people, based on representative data, for example political parties or companies. Although they have bottom-up elements, a great deal of decision-making is done in a top-down manner. However, as systems become more complex, they will require more bottom-up elements or they will otherwise perform poorly (see Sec. \ref{trafficlight}) or even destabilize over time \citep{HelbingGloballyNetworkedRisks}.

Digital technologies are now enabling more flexible ways of organization. In particular, people may use social media platforms to organize ``projects'' in a bottom-up way. In principle everyone could do this, given the required technical and social skill sets. This development could eventually turn consumers into ``prosumers'',\footnote{see http://en.wikipedia.org/wiki/Prosumer} i.e. consumers who are co-creating products they buy (and sell) \citep{ProsumerPower}). 3D printer technology, for example, will enable local production by small teams or individuals who may sell their products to friends, colleagues or the rest of the world. Rather than just specifying the color and individual features of a car when ordering, we may soon design some of its components and commission their production. One may even set up a team of designers, engineers, marketing people, and other specialists to design an own car with components produced by other companies or with new components commissioned from home. That is, old-style factories and socionomies based on collaborative projects are expected to work hand in hand.

To a certain extent, ``projects'' of the above kind are already in existence today, for example open source software projects. Many such projects are driven by volunteers or employees of companies, who rely on open source components and want to get their required features implemented. In any case, the development is bottom-up and open. The ``open source ecosystem'' is based on a number of ingredients such as ``viral'' open source licenses, which ensure that those using open source code in their own software will also have to make it available to others. Therefore, software licenses (such as the GNU General Public License) reward other-regarding behavior as it characterizes the ``homo socialis''. In the context of open source development, the GitHub platform\footnote{\url{https://github.com}} has recently become very popular among software developers. The platform indicates who has contributed how much, thereby creating incentives for contributing. Thus, everyone can benefit from a growing set of open source software.

The digital economy will open up infinite opportunities for new products. We might even say that the information age enables infinite dimensions of creativity. This has very interesting implications for the resulting market structures. Today, we have a few core businesses and some peripheral business activities. In the future, however, one may expect that projects and peripheral products will dominate market activities.\footnote{One may imagine this as the core of a sphere as compared to its surface area. Note, however, that the relationship between the surface area $A$ of an n-dimensional sphere and its volume $V$ is $V = rA/n$, see \url{http://en.wikipedia.org/wiki/Sphere}. That is, the higher the dimensionality of a market, the more happens in the peripheral (surface) area.} This means that many more products will be individually customized. Note also, that projects imply new forms of work and employment.\footnote{Amazon Mechanical Turk, for example, is a platform that matches tasks and workforce.} But if suitable institutions exist, projects could create a large number of jobs in the future. Therefore, socionomies could not only unleash an age of creativity, but also provide novel ways to overcome the current unemployment misery.

It is likely that, over time, companies, political parties, and other established institutions will be complemented by ``projects'' as more flexible form of organization. Then, the future role of entrepreneurs will be to set up and coordinate such projects, and organize the necessary supports. Once completed, a project would terminate, and the previously involved people would look for new projects to coordinate or participate in. In this way, a  ``participatory market society'' would emerge.\footnote{Note that ``cooperatives'', which are a common form of business organization in Switzerland and elsewhere, fit a participatory market society well, but ``projects'' are expected to be more short-lived and flexible.}  Everyone could be both a coordinator and a participant of several projects, which provides opportunities to influence issues one cares about. The fluent, project-based organization of socio-economic activities might also be a good solution to the Peter Principle \citep{PeterPrinciple}, according to which people get promoted until they end up in a position, which overstrains their abilities.

``Projects'' will also imply more self-determined and exciting work. To unleash creativity, decision-making needs to take place differently from the way it works today. Currently, many decisions are majority-based or top-down decisions. In the first instance, a majority decision is a compromise, which is often disappointing to all (the ``common denominator''). In the second case, someone  imposes decisions on others, which are potentially against their preferences. This creates advantages for some and disadvantages for others. However, it should be remembered that diversity is the main driving force of innovation in evolutionary systems \citep{page,Inno,Inno2,Inno3,Helbing05b}. It would, therefore, be better to allow projects to find their own rules.\footnote{Future ICT systems will support and simplify coordination between people with different interests and backgrounds. ``Inter-cultural adapters'' will help to overcome the need for everyone to agree on the same rules and principles. Reputation systems will promote a proper quality (see Sec. \ref{ReputationSystems}).} People could then easily find projects and socio-economic environments that fit their individual preferences, interests, and needs.




\section{Conclusions and Outlook}

Globalization and technological revolutions have created  new levels of interdependencies, interconnectedness and complexities, which have the potential to  destabilize our techno-socio-economic-environmental system(s) on a global scale \citep{HelbingGloballyNetworkedRisks}. It is therefore critical that, new approaches are found to stabilize global networks and counter systemic instability. 

The current trend seems to point towards ``surveillance societies'' with strong punitive elements (``punitive societies''). However, such a top-down approach would endanger privacy, socio-diversity and innovation. It also implies major risks that personal data will be sooner or later misused and that a transition to a totalitarian state can occur \citep{GoogleAsGod}. Furthermore, recent findings suggest that crime cannot be eliminated by large fines \citep{inspectiongame}, contrary to what classical rational choice models suggest, and there is little evidence that surveillance by CCTV cameras can reduce crime in a systematic and statistically significant way \citep{CCTV}.\footnote{UK has surprisingly high crime rates in Europe, given the high coverage with CCTV cameras. A recent terror attack in London intentionally committed below a CCTV camera also questions whether surveillance can be an efficient strategy to stabilize society.} Trust, when substantiated, seems to be a more sustainable basis of a stable society. In fact, another recent trend points towards ``reputation societies'', but the underlying reputation systems should be user-centric, based on multiple criteria, and run in a decentralized way \citep{Zhou,Mundinger}. 

I expect that decentralized self-organization and self-regulation approaches can deliver solutions for complex dynamical systems that are far superior to our existing way of designing and operating systems. However, self-regulation does not mean that one can choose the rules as one likes. In fact, as could be seen for the urban traffic light control example, self-regulation on the component level generally does not work, but it must be other-regarding (which requires a certain amount of trustable information exchange).\footnote{Also note that the most efficient way of self-regulation may contain hierarchical elements, combining top-down and bottom-up elements in a favorable way. So, at least in the next decades, I do not expect reputation systems to replace top-down elements of social organization, but rather to complement them. Also, ``qualified money'' would exist next to our current kind of money. However, I expect the relative proportion of self-regulation (bottom-up organization) to grow substantially over time.}

To be successful, societies must be able to resolve ``social dilemma situations'', in particular, free-riding must be efficiently contained. In social dilemma situations, cooperative behavior would be favorable for everyone, but exploiting others creates even higher payoffs. Such situations often lead to ``tragedies of the commons'', where environmental pollution, overfishing, global warming, free-riding, tax evasion, and the exploitation of our social benefit systems are typical results. Problems like these are expected in particular for the ``homo economicus'', who takes optimally self-regarding decisions. In spite of this, the ``homo economicus'' is still a foundation of mainstream economic thinking and many policies today.

As non-linear interaction and network effects become more and more important for today's decision-making \citep{schweitzer2009,PositiveLinking}, a new approach is needed. The approach of an other-regarding ``homo socialis'' now seems to be a more promising theoretical starting point than the independently deciding ``homo economicus''. Like the ``homo economicus'', the ``homo socialis'' results from the merciless forces of natural selection, but it can overcome ``tragedies of the commons'' and, therefore, achieve higher payoffs.

To understand this, it is important to study economic systems from the viewpoint of complexity theory. We have seen that 

the combination of non-linear interactions with randomness (such as ``mutations'') can produce counter-intuitive emergent phenomena. Complexity theory could also provide a new perspective on socio-economic systems more generally \citep{HelbingRethinkingEconomics}. The policy implications can be quite different from those for the ``homo economicus''. Therefore, one can expect novel insights and fundamentally new solutions to old problems, including (over-)regulation, innovation, financial crises, unemployment, and sustainability.

As Albert Einstein said: ``We cannot solve our problems with the same kind of thinking that created them''. In the past, there has been a steady struggle between economic activities and social motives, and between bottom-up market organization and political top-down regulation. Such regulation serves to avoid exploitation, ``tragedies of the commons'', and market failures, whenever the self-organization of the ``homo economicus'' does not lead to desirable outcomes. The ``homo socialis'' bridges the gap between these apparently incompatible sides. However, as the interdependent decisions of other-regarding, ``networked minds'' cause a complex system dynamics and different outcomes on the macro-level, a new scientific discipline is needed, called ``economics 2.0'' or  ``socionomics''. This research field will have to study the implications of the ``homo socialis'' and supportive institutional settings, particularly successful principles of self-regulation.




The new conceptual framework may be characterized as follows:

\begin{enumerate}

\item The ``homo socialis'' takes self-determined, but responsible decisions, caring about the impact on others. While this fits well with Kant's moral imperative\footnote{see also the somewhat related work of \cite{kantian}} and with the values promoted by many religions, it does not need to be based on ethical or religious grounds \citep{NelsonEconomicsAsReligion}. The concept is built on scientific insights showing that other-regarding behavior helps to coordinate between incompatible interests, leading to better individual outcomes over a long enough time horizon.

\item As the ``homo socialis'' is able to get more payoff under conditions that are highly competitive, the concept of participatory market societies is not based on an idealistic approach. There is also much empirical evidence in our digital economy for this.

\item The approach discussed in this paper promotes social order and social welfare, but also sustainable prosperity,
individual freedom, diversity and pluralism. It is very different from the concept of a central regulator discussed in Sec. \ref{trafficlight}. In particular, it does not impose a solution, which it considered best for everyone, on others, be it equality or something else.

\item The new approach rather builds on suitable principles of self-organization and self-regulation based on decentralized decisions. This will allow one to cut down today's overregulation, thereby creating larger individual freedoms.

\item Efficient ICT-enabled reputation systems and ``qualified money'' might support such self-regulation, if they are open, user-centric, participatory, and based on sufficiently many criteria. They could create an institutional setting allowing the ``homo socialis'' and the ``homo economicus'' to coexist.\footnote{Note that both the ``homo economicus'' and the ``homo socialis'' are needed for a diverse and pluralistic society to function well. For example, the ``wisdom of crowds'' should work well for independent decision-makers (the ``homo economicus''), while it might be undermined by social influence (as it is characteristic for the ``homo socialis'') \citep{Lorenz11}. However, the ``homo socialis'' seems to be better in dealing with social dilemma situations. So, both kinds of behavior have their advantages and disadvantages, depending on the respective context.} ``Qualified money'' would generalize established principles of value exchange at stock markets to units of money. In particular, it would consider that information is a precondition for markets to function, while today's money is memory-less, thereby cutting away important information that may allow financial systems to be more adaptive and resilient. I expect the best solution to be a market system in which conventional and qualified money would exist in parallel with suitable conversion rules.\footnote{The reputation system will be described in more detail elsewhere. It may allow users to determine themselves whether they want to share the reputation information about them with others and with whom. It should also establish a suitable balance between traceable and anonymous (information or financial) exchange, thereby limiting misuse of anonymity and stimulating responsible action on the one hand, while providing individual freedoms on the other hand. Finally, the reputation system could include the feature of ``forgetting'' and ``forgiving''.}

\end{enumerate}


The Web 2.0 is a major driver of the transition from the conventional market economy (``economy 1.0'') to a ``participatory market society'' (``economy 2.0''). The change is largely fueled by social media platforms, 3D printers enabling local production, and ``Big Data'' celebrated as the ``Oil of the 21st Century''. It is also driven by an increasing level of complexity, which can only be mastered by a higher level of decentralization, as implied by the concept of self-regulation\footnote{Andrew Haldane at the Bank of England puts it like this: ``Modern finance is complex, perhaps too complex. Regulation of modern finance is complex, almost certainly too complex. That configuration spells trouble. As you do not fight fire with fire, you do not fight complexity with complexity. Because complexity generates uncertainty, not risk, it requires a regulatory response grounded in simplicity, not complexity.'' \citep{Haldane12}}.

This trend towards more decentralization is already visible in the way the Internet is organized, the way smart grids are now being run, and the way modern traffic systems will be managed. For socio-economic systems, the trend towards decentralization implies the participation of more people in social, economic and political affairs. A historical perspective also suggests that more and more people gain influence on decision-making processes as centuries progress.\footnote{This may be seen as a consequence of time scale separation, which is a precondition for power in hierarchical systems \citep{HelbingGloballyNetworkedRisks,HelbingSSO}. One may imagine the result of this process to be a ``Swiss basic democracy plus'', enabled by future ICT systems.}

In fact, the digital revolution is already reshaping our economy \citep{Pyka,Leydesdorff,Economics2.0}. Social media have become highly successful and influential. App Stores have created an ecosystem of millions of apps. Even when provided for free, they can create considerable economic value.  Furthermore, the ``sharing economy'' is currently generating growth rates of 20\%.  So, participatory market societies have great potential. 

Humanity has now the opportunity to enter an era of creativity and prosperity built on social principles and enabled by modern ICT systems. It just takes a proper understanding of socionomic systems to determine and establish suitable institutions such as open, user-centric, multi-criteria reputation systems. In the past, humans have created institutions such as markets, roads, courts, museums, schools, libraries, and universities to benefit our society and economy. It is now time to create the right institutions for the information society, supporting socio-economic life in the 21st century. If the right decisions are made, these institutions will unleash the creativity of our ``networked minds'', thereby opening the door to an age of sustainable prosperity.

\subsection*{Acknowledgments}

The author would like to thank Stefano Balietti, Anna Carbone, Thomas Chadefaux, Andreas Diekmann, Tobias Kuhn, Matthias Leiss, and Heinrich Nax for inspiring discussions and critical comments. Special thanks to Christian Waloszek, Maria Kaninia and Stefan L\"ammer for preparing the figures and for many useful inputs. 

The author also acknowledges the support of the European Commission through the FET Flagship Pilot Project FuturICT (Grant No. 284709) and the ERC Advanced Investigator Grant ``Momentum'' (Grant No. 324247).


\nocite{Hobbes,SmithMoralSentiments,SmithWealthOfNations,Arthur89,Arthur99,Selten98,Stiglitz11,Krugman90,Kahneman79,Kahneman84,Kahneman96,Tversky92,Gigerenzer99,BFrey,Fehr03,Fehr99,GintisBoundsOfReason,Witt98}

\bibliographystyle{aer} 

\bibliography{Socionomy_References_8}

\end{document}